\documentclass[runningheads]{cl2emult}
\begin{document}

\def\C{{\bf C}}
\def\H{{\cal H}}
\def\O{{\cal O}}
\def\A{{\cal A}}
\def\B{{\cal B}}
\def\PH{{\cal PH}}
\def\shalf{\hbox{${\textstyle{1\over 2}}$}}
\def\op#1{\hbox{\sf #1}}
\def\ihbar{\hbox{${\textstyle{i\over\hbar}}$}}
\def\Mhbar{\hbox{${\textstyle{M\over\hbar}}$}}
\def\hbarM{\hbox{${\textstyle{\hbar\over M}}$}}
\def\reals{\hbox{${\bf R}$}}
\def\b#1{\bar{#1}} 
\def\t#1{\tilde{#1}}

      \title*{States, Symmetries and Superselection}
    \toctitle{States, Symmetries and Superselection}
\titlerunning{States, Symmetries and Superselection}

\author{Domenico Giulini\inst{}}
\authorrunning{Domenico Giulini}
\institute{Theoretische Physik, Universit\"at Z\"urich,\\
           Winterthurerstrasse~190, CH-8057 Z\"urich, 
           Switzerland}

\maketitle

\begin{abstract}
The process of dynamical decoherence may cause apparent 
superselection rules, which are sometimes called `environmentally 
induced' or `soft'. A natural question is whether such dynamical 
processes are eventually also responsible for at least some of 
the superselection rules which are usually presented as 
fundamentally rooted in the kinematical structure of the theory 
(so called `hard' superselection rules). With this question in 
mind, I re-investigate two well known examples where 
superselection rules are usually argued to rigorously exist within 
the given mathematical framework. These are (1)~the Bargmann 
superselection rule for the total mass in Galilei invariant quantum 
mechanics and (2)~the charge superselection rule in quantum 
electrodynamics. I argue that, for various reasons, the kinematical 
arguments usually given are not physically convincing unless they 
are based on an underlying dynamical process. 
\end{abstract}
\section{Introduction}
Fundamental to the concept of dynamical decoherence is the notion of 
`delocalization' \cite{Joos}\cite{GJKKSZ}. The intuitive idea 
behind this is that through some dynamical process certain state 
characteristics (`phase relations'), which were locally accessible at 
one time, cease to be locally accessible in the course of the dynamical
evolution. Hence locally certain superpositions cannot be 
verified anymore and an apparent obstruction to the superposition 
principle results. Such mechanisms can therefore be considered 
responsible for so-called soft superselection rules, like that of 
molecular chirality~\cite{Wightman1}. They are called `soft', because 
they only hold with respect to the limited class of local observables
and are hence of approximate validity. But there are also `hard'
superselection rules, which are usually presented as rigorous 
mathematical results within the kinematical framework of the theory.
Such presentations seem to suggest that there is no room left for a dynamical 
interpretation, and that hence these two notions of superselection
rules are really distinct. Here I wish to argue that at least 
{\it some} of the existing proofs for `hard' superselection rules 
give a false impression, and that quite to the contrary they actually 
need some dynamical input in order to be physically
convincing. We will 
look at the case of Bargmann's superselection rule for total mass
in ordinary quantum mechanics and that of charge in quantum 
electrodynamics (QED). The discussion of the latter will be heuristic 
insofar as we will pretend that QED is nothing but quantum mechanics 
(in the Schr\"odinger representation) of the infinite-dimensional 
(constrained) Hamiltonian system given by classical
electrodynamics.

Crucial to the ideas presented here is of course that `delocalized'
does not at all mean `destroyed', and that hence the loss of quantum 
coherence is only an {\it apparent} one. This distinction might 
be considered irrelevant FAPP\footnote{{\bf F}or {\bf A}ll 
{\bf P}ractical {\bf P}urposes.},
but it is important in attempts to understand apparent losses of 
quantum coherence {\it within} the standard dynamical framework of
quantum mechanics.

As used here, the term `local' usually refers to locality in 
the (classical) configuration space $Q$ of the system, where we 
think of quantum states in the Schr\"odinger representation, i.e., as 
$L^2$-functions on $Q$. Every parametrization of $Q$ then defines
a partition into `degrees of freedom'. Locality in $Q$ is a more general
concept than locality in ordinary physical space, although the latter
forms a particular and physically important special case. Moreover, on a
slightly more abstract level, one realizes that the most general
description of why decoherence appears to occur is that only a
{\it restricted} set of so-called physical observables are at ones
disposal, and that  {\it with respect to those} the relevant `phase
relations' {\it apparently} fade out of existence. It is sometimes
convenient to express this by saying that decoherence occurs only
with respect (or relative) to a `choice' of observables.  Clearly this
`choice' is not meant to be completely free, since it has to be
compatible with the dynamical laws and the physically realizable
couplings (compare \cite{Joos}). But to fully control those is 
a formidable task -- to put it mildly -- and a careful a priori 
characterization of observables seems therefore almost always 
inescapable. In this respect the situation bears certain 
similarities to that of `relevant' and `irrelevant' degrees of 
freedom in statistical mechanics.

\section{Hilbert Spaces and Observables}
The mathematical modelling of a physical system 
involves a specification of a space of states and a space
of observables. In quantum theory this is usually done with 
the help of the underlying classical theory. States are then 
identified with the space of rays in the Hilbert space
$\H=L^2(Q)$, which we denote by $\PH$ (projective Hilbert space)
and observables are certain operators on $\H$. 
I am not aware of any generally valid criteria by which one 
might fully characterize sets of operators on $\H$ `as physical 
observables'. But there are certain mathematical structures which 
seem physically admissible and also natural, which, once imposed, 
allow to make some general statements about the set of physical 
observables.
 
Self-Adjointness is a generally accepted criterion, and without 
loss of generality one can also restrict to bounded operators.
Although only certain combinations of addition and multiplication 
preserve self-adjointness, it is mathematically more convenient to 
consider the whole algebra generated (in a sense made more precise 
below) by this set. This algebra is then called the algebra of physical
observables, $\O$, although only its self-adjoint elements actually  
correspond to observables. Moreover, since physically the matrix 
elements are the relevant quantities, it is natural to require that 
a sequence of operators converges if and only if (henceforth: iff)
all the matrix elements converge; 
in technical terms, the algebra should be (topologically) closed in
the weak operator topology, i.e., be a von Neumann algebra.
An extremely useful fact is that weak closures of algebras can be 
characterized in purely algebraic terms. This works as follows: Let
$\B(\H)$ denote the algebra of bounded operators on 
$\H$. If $\A\subseteq\B(\H)$ is any subset, then 
$\A':=\{B\in \B(\H)\,:\, AB=BA,\ \forall A\in\A\}$ is called the
commutant of $\A$. Iterating this procedure leads to $\A''$, of which 
the following is true: (i)~$\A''$ is a von Neumann algebra, (ii)~it is 
the smallest von Neumann algebra containing $\A$. In this sense one says 
that $\A''$ is the von Neumann algebra {\it generated} by $\A$. In particular,
if $\A$ was already a von Neumann algebra, it must satisfy $\A=\A''$.
Note also that, by definition of the commutant, an inclusion of the
form $\A\subseteq \B$ implies $\B'\subseteq \A'$. 

Recall that the center, $\O^c$, of $\O$ consists 
of those elements in $\O$ which also lie in the commutant $\O'$, i.e., 
commute with all elements in $\O$; hence $\O^c:=\O\cap\O'$. 
Complex multiples of the identity operator are trivially 
contained in $\O^c$ and any other ones are called superselection 
operators. If $\O^c$ contains a projection operator onto a 
subspace $\H'\subset\H$, then $\O$ must leave $\H'$ invariant 
and $\H'$ is said to reduce $\O$. In this case the theory is said to 
possess superselection rules. In the opposite case, $\O$ acts 
irreducibly on $\H$ and no superselection rules exist.

We now list some simple and general criteria which have been 
suggested in the literature to characterize $\O$. These criteria 
will involve $\O$ and $\O'$ and hence do not only concern 
the abstract algebraic object $\O$, but also $\H$ on which $\O'$ 
depends. Hence they can either be read as condition on 
$\O\subseteq\B(\H)$ {\it given} $\H$, or as certain `matching 
conditions' between the two mathematical objects representing 
physical observables on one hand and states on the other.

\medskip\noindent
$\bullet$ In 1932, von Neumann \cite{vNeumann} proposed to identify 
$\O=B(\H)$. Then $\O'=\{\alpha{\bf 1},\ \alpha\in\C\}$ and hence no
superselection rules exist.

\medskip\noindent
$\bullet$ In 1952, Wick, Wightman, and Wigner \cite{WWW} considered 
for the first time the possibility that $\O$ might be strictly 
smaller than $B(\H)$ and not act irreducibly on $\H$, so that  
$\H=\oplus_i\H_i$, where each $\H_i$ reduces $\O$. The projectors 
onto the $\H_i$ obviously lie in $\O^c$ and hence are superselection 
operators. Now only the rays in $\bigcup_i\H_i$ define pure states. 
This can be generalized to direct integrals.

\medskip\noindent
$\bullet$ In 1960, Jauch \cite{Jauch1} properly formulated 
Dirac's condition, that there should exist at least one 
`complete set of commuting observables', which in this formulation 
makes sense only for operators with discrete spectra. The generally 
valid formulation is, that $\O$ should contain a von Neumann 
subalgebra $\A\subseteq\O$ which is {\it maximal abelian}. Being 
abelian clearly means that $\A\subseteq\A'$ and being maximal 
means that everything that commutes with $\A$ is already contained 
in it, i.e., $\A'\subseteq\A$. Taken together, maximal abelian
is equivalent to $\A=\A'$. This condition can be read as saying 
that $\O$ cannot be too small, since it must accommodate an abelian 
subalgebra that is
maximal.\footnote{Without further qualification the term
`maximal' always means `maximal {\it in} $\B(\H)$'. 
In contrast, $\A\subseteq\O$  is said to be  maximal abelian 
{\it in} $\O$ if $\A=\A'\cap\O$. 
It would be pointless to require maximality in $\O$, since such
subalgebras always exist (by Zorn's Lemma). Obviously maximality 
in $\B(\H)$ implies maximality in $\O$. But in the general case the
converse is true iff the Dirac-Jauch condition 
$\O'\subseteq \O$ is met. Proof: 
Suppose (a)~$\O'\subseteq\O$,
        (b)~$\O'\subseteq\A'$ (obvious from $\A\subseteq\O$), 
    and (c)~$\A=\A'\cap\O$. 
Then for $Z\in\O'$ (a), (b) and (c) immediately imply $Z\in\A$, 
hence $\O'\subseteq\A$ or $\A'\subseteq \O''=\O$. 
Then (c) is equivalent to $\A=\A'$.}
Now, the point is that the existence of a maximal abelian 
$\A\subseteq\O$ can be equivalently expressed just in terms of $\O$
and $\O'$, namely by saying that everything that commutes with $\O$
is contained in $\O$, that is,  $\O'\subseteq \O$ (Dirac-Jauch
condition).
Necessity of this condition is readily seen, since $\A\subseteq\O$
implies
$\O'\subseteq\A'=\A\subseteq \O$. The converse was shown by
Jauch~\cite{Jauch1}. Note also that $\O^c:=\O\cap\O'$ implies that 
$\O'\subseteq\O$ can be rewritten as $\O^c=\O'$, which essentially 
says that the center of $\O$ already exhausts the set of all
those elements in $\B(\H)$ commuting with $\O$.

\medskip\noindent
$\bullet$ In 1961, Jauch and Misra \cite{Jauch2} discussed the 
relation between superselection rules and gauge symmetries
(supersymmetries in their language). Since by definition gauge 
symmetries commute with physical observables, they are generated 
by the unitary elements of $\O'$. But since $\O$ is a von~Neumann
algebra, we have $\O=\O''$. Therefore $\O'\subseteq\O$ is equivalent
to $\O'\subseteq\O''$, i.e. that $\O'$ is abelian. The Dirac-Jauch
condition is therefore equivalent to the requirement that 
gauge groups should be abelian.\footnote{Note that the statement 
is not that gauge symmetries need {\it a priori} be abelian. 
For example, the permutation group for $n$ particles ($n>2$) 
is perfectly legitimate to start with. However, the unreduced 
$n$-particle Hilbert space is now definitely too big in the sense 
that it contains higher-dimensional subspaces in which each 
ray defines the {\it same} pure state. Such a redundancy is against 
the Dirac-Jauch condition, which in particular implies that $\O$ should 
separate the rays in the space of physical states. In this case the 
Dirac-Jauch condition is met by truncating $\H$ so as to 
leave only one ray per pure state. Thereby the original gauge group 
is broken down to a residual one which is abelian. 
See \cite{Giulini1}.}

\section{Superselection Rules via Symmetry Requirements}
The requirement that a certain group must act on the set of all
physical states is often the (kinematical) source of superselection
rules. Here I wish to explain the structure of this argument.

Note first that in quantum mechanics we identify the states of a 
closed system with rays and not with vectors which represent them 
(in a redundant fashion).
It is therefore not necessary to require that a symmetry group $G$ 
acts on the Hilbert space $\H$, but rather it is sufficient that 
it acts on $\PH$, the space of rays, via so-called ray-representations.
Mathematically this is a non-trivial relaxation since not every 
ray-representation of a symmetry group $G$ (i.e. preserving the ray products) 
lifts to a unitary action of $G$ on $\H$. What may go wrong is not that 
for a given $g\in G$ we cannot find a unitary (or anti-unitary) 
operator ${\op U}_g$ on $\H$; that is assured by Wigner's 
theorem (see \cite{Bargmann} for a proof).  
Rather, what may fail to be possible is that we can choose the 
${\op U}_g$'s in such a way that we have an {\it action}, i.e., that 
${\op U_{g_1}}{\op U_{g_2}}=U_{g_1g_2}$. As is well known, this is 
precisely what happens for the implementation of the Galilei group in 
ordinary quantum mechanics. Without the admission of ray 
representations we would not be able to say that ordinary quantum 
mechanics is Galilei invariant.   

To be more precise, to have a ray-representation means that for each 
$g\in G$ there is a unitary\footnote{For simplicity we ignore 
anti-unitary transformations. They cannot arise if, for example, 
$G$ is connected.} transformation ${\op U}_g$ which, instead 
of the usual representation property, are only required
to satisfy the weaker condition
\begin{equation}
{\op U}_{g_1}{\op U}_{g_2}=\exp(i\xi(g_1,g_2))\,{\op U}_{g_1g_2},
\label{ray}
\end{equation}
for some function $\xi:G\times G\rightarrow\reals$, called multiplier
exponent, satisfying\footnote{The following conditions might seem 
a little too strong, since it would be sufficient to require 
the equalities in (2) and (\ref{cocycle}) only mod 
$2\pi$; this also applies to (\ref{redef}). But for our application in 
section~4 it is more convenient to work with strict equalities, 
which in fact implies no loss of generality.}
\begin{eqnarray}
\xi(1,g)=\xi(g,1)&=&0, \\
\label{zero}
\xi(g_1,g_2)-\xi(g_1,g_2g_3)+\xi(g_1g_2,g_3)-\xi(g_2,g_3)&=&0.
\label{cocycle}
\end{eqnarray}
The second of these conditions is a direct consequence of 
associativity: ${\op U_{g_1}}({\op U_{g_2}}{\op U_{g_3}})=
({\op U_{g_1}}{\op U_{g_2}}){\op U_{g_3}}$.
Obviously these maps project to an action of $G$ on $\PH$. Any other 
lift of this action on $\PH$ onto $\H$ is given by a redefinition
${\op U}_g\rightarrow {\op U}'_g:=\exp(i\gamma(g)){\op U}_g$, for some 
function $\gamma:G\rightarrow\reals$ with $\gamma(1)=0$, 
resulting in new multiplier exponents
\begin{equation}
\xi'(g_1,g_2)=\xi(g_1,g_2) +\gamma(g_1)-\gamma(g_1g_2)+\gamma(g_2),
\label{redef}
\end{equation}
which again satisfy (2) and (\ref{cocycle}). The ray 
representations ${\op U}$ and
${\op U'}$ are then said to be equivalent, since the projected actions on
$\PH$ are the same. We shall also say that two multiplier exponents
$\xi,\xi'$ are equivalent if they satisfy (\ref{redef}) for some $\gamma$.

We shall now see how the existence of inequivalent multiplier exponents 
together with the requirement that the group should 
act on the space of physical states, may clash with the 
superposition principle and thus give rise to superselection rules.  
For this we start from two Hilbert spaces $\H'$ and $\H''$ and 
actions of a symmetry group $G$ on $\PH'$ and $\PH''$, i.e., ray 
representations ${\op U'}$ and ${\op U''}$ on $\H'$ and $\H''$ up to 
equivalences (\ref{redef}). We consider $\H=\H'\oplus\H''$ and ask under 
what conditions does there exist an action of $G$ on $\PH$ which 
restricts to the given actions on the subsets $\PH'$ and $\PH''$. 
Equivalently: when is ${\op U}={\op U'}\oplus{\op U''}$ a ray 
representation of $G$ on $\H$ for some choice of ray-representations 
${\op U}'$ and ${\op U}''$ within their equivalence class? To answer 
this question, we consider 
\begin{eqnarray}
{\op U}_{g_1}{\op U}_{g_2}
&=& ({\op U}'_{g_1}\oplus{\op U}''_{g_1})
    ({\op U}'_{g_2}\oplus{\op U}''_{g_2}) \nonumber\\
&=& \exp(i\xi'(g_1,g_2)){\op U}'_{g_1g_2}
    \oplus\exp(\xi''(g_1,g_2)){\op U}''_{g_1g_2}
\label{sum}
\end{eqnarray}
and note that this can be written in the form (\ref{ray}), for some choice 
of $\xi',\xi''$ within their equivalence class, iff the phase factors can 
be made to coincide, that is, iff $\xi'$ and $\xi''$ are equivalent.
This shows that there exists a ray-representation on
$\H$ which restricts to the given equivalence classes of given ray
representations on $\H'$ and $\H'$, iff the multiplier exponents
of the latter are equivalent. Hence, if the multiplier 
exponents $\xi'$ and $\xi''$ are not equivalent, the action of $G$
cannot be extended beyond the disjoint union $\PH'\cup\PH''$.
Conversely, {\it if} we require that the space of physical states 
must support an action of $G$ then non-trivial superpositions 
of states in $\H'$ and $\H''$ must be excluded from the space of 
(pure) physical states. 

This argument shows that if we insist of implementing $G$ 
as symmetry group, superselection rules are sometimes 
unavoidable. A formal trick to avoid them would be not to 
require $G$, but a slightly larger group, $\b G$, to act on the 
space of physical states. $\b G$ is chosen to be the group 
whose elements we label by $(\theta,g)$, where $\theta\in\reals$, 
and the multiplication law is 
\begin{equation}
{\b g}_1{\b g}_2 = (\theta_1,g_1)(\theta_2,g_2)
                 = (\theta_1+\theta_2+\xi(g_1,g_2),g_1g_2).
\label{extension}
\end{equation}
It is easy to check that the elements of the form $(\theta, 1)$ 
lie in the center of $\b G$ and form a normal subgroup 
$\cong\reals$ which we call $Z$. Hence $\b G/Z=G$ but $G$ 
need not be a subgroup of $\b G$. $\b G$ is a central $\reals$ 
extension\footnote{Had we defined the multiplier exponents mod 
$2\pi$ (compare footnote 5) then we would have obtained a $U(1)$ 
extension, which would suffice so far. But in the next section 
we will definitively need the $\reals$ extension as symmetry group  
of the extended classical model discussed there.}
of $G$~ (see e.g. \cite{Raghunathan}). 
Now a ray-representation ${\op U}$ 
of $G$ on $\H$ defines a proper representation $U$ of
$\b G$ on $\H$ by setting $U_{(\theta,g)}:=\exp(i\theta){\op U_g}$.
Then $\b G$ is properly represented on $\H'$ and $\H''$ and hence 
also on $\H=\H'\oplus\H''$. The above phenomenon is mirrored here 
by the fact that $Z$ acts trivially on $\PH'$ and $\PH''$
but non-trivially on $\PH$, and the superselection structure
comes about by requiring physical states to be fixed points of 
$Z$'s action.

\section{Test Case: Bargmann's Superselection Rule}
An often mentioned textbook example where a particular 
implementation of a symmetry group allegedly clashes with the 
superposition principle, such that a superselection rule results, 
is Galilei invariant quantum mechanics (e.g. \cite{Galindo}; see
also Wightman's review \cite{Wightman2}). We will discuss this 
example in detail for the general multi-particle case. (Textbook 
discussions usually restrict to one particle, which, due to Galilei
invariance, must necessarily be free.) It will serve as a test case to 
illustrate the argument of the previous chapter and also to formulate 
my critique. Its physical significance is limited by the fact that the 
particular feature of the Galilei group that is responsible for the 
existence of the mass superselection rule ceases to exist if we replace 
the Galilei group by the Poincar\'e group (i.e. it is unstable under 
`deformations'). But this is not important for my argument.
\footnote{In General Relativity, where the total mass can be 
expressed as a surface integral at `infinity', the issue of mass 
superselection comes up again; see e.g. \cite{GKZ}.}

Let now $G$ be the Galilei group, an element of which is parameterized
by $(R,\vec v,\vec a,b)$, with $R$ a rotation matrix in $SO(3)$,
$\vec v$ the boost velocity, $\vec a$ the spatial translation,
and $b$ the time translation. Its laws of multiplication and inversion 
are respectively given by 
\begin{eqnarray}
g_1g_2 &=& (R_1,\vec v_1,\vec a_1,b_1)(R_2,\vec v_2,\vec a_2,b_2)\nonumber\\
       &=& (R_1R_2\,,\,\vec v_1+R_1\cdot\vec v_2\,,\,a_1+R_1\cdot\vec a_2+
            \vec v_1b_2\,,\,b_1+b_2),
\label{mult}\\
g^{-1}&=& (R,\vec v,\vec a,b)^{-1}=(R^{-1},\,-R^{-1}\cdot\vec v\,,\, 
-R^{-1}\cdot(\vec a-\vec vb)\,,\,-b).
\label{inv}
\end{eqnarray} 
We consider the Schr\"odinger
equation for a system of $n$ particles of positions $\vec x_i$,
masses $m_i$, mutual distances $r_{ij}:=\Vert\vec x_i-\vec x_j\Vert$
which interact via a Galilei-invariant potential $V(\{r_{ij}\})$,
so that the Hamilton operator becomes
${\op H}=-\hbar^2\sum_i\frac{\Delta_i}{2m_i}+V$. The Hilbert space
is $\H=L^2(\reals^{3n},d^3\vec x_1\cdots d^3\vec x_n)$.

$G$ acts on the space 
$\{\hbox{configurations}\}\times\{\hbox{times}\}\cong\reals^{3n+1}$ 
as follows: Let $g=(R,\vec v,\vec a,b)$, then 
$g(\{\vec x_i\},t):=(\{R\cdot\vec x_i+\vec vt+\vec a\}\,,\,t+b)$.
Hence $G$ has the obvious left action on complex-valued functions on 
$\reals^{3n+1}$: $(g,\psi)\rightarrow\psi\circ g^{-1}$. However, these
transformations do {\it not} map solutions of the Schr\"odinger equations
into solutions. But as is well known, this can be achieved by introducing 
an $\reals^{3n+1}$-dependent phase factor (see e.g. \cite{Giulini2}). 
We set $M=\sum_i m_i$ for the total mass and 
$\vec r_c=\frac{1}{M}\sum_im_i\vec x_i$ for the center-of-mass. 
Then the modified transformation, ${\op T}_g$, which maps solutions 
(i.e. curves in $\H$) to solutions, is given by
\begin{equation}
{\op T}_g\psi(\{\vec x_i\},t):=
\exp\left(\ihbar M[\vec v\cdot(\vec r_c-\vec a)
-\shalf\vec v^2(t-b)]\right)\,\psi(g^{-1}(\{\vec x_i\}, t)).
\label{trans}
\end{equation}
However, due to the modification, these transformations have lost the 
property to define an action of $G$, that is, we do {\it not} have 
${\op T}_{g_1}\circ {\op T}_{g_2}={\op T}_{g_1g_2}$. Rather, a straightforward 
calculation using (\ref{mult}) and (\ref{inv}) leads to 
\begin{equation}
{\op T}_{g_1}\circ{\op T_{g_2}}
=\exp(i\xi(g_1,g_2))\,{\op T}_{g_1g_2},
\label{raytrans}
\end{equation}
with non-trivial multiplier exponent
\begin{equation}
\xi(g_1,g_2)=
\Mhbar({\vec v}_1\cdot R_1\cdot{\vec a}_2+\shalf{\vec v}_1^2b_2).
\label{Gphase}
\end{equation} 
Although each ${\op T}_g$ is a mapping of {\it curves} in 
$\H$, it also defines a unitary transformation on $\H$ itself. 
This is so because the equations of motion define a bijection 
between solution curves and initial conditions at, say, $t=0$,
which allows to translate the map ${\op T}_g$ into a unitary 
map on $\H$, which we call ${\op U}_g$. It is given by  
\begin{equation}
{\op U}_g\psi(\{\vec x_i\})=
\exp\left(\ihbar M[\vec v\cdot(\vec r_c-\vec a)
+\shalf\vec v^2b]\right)\,
\exp(\ihbar{\op H}b)\psi(\{R^{-1}(\vec x_i-\vec a+\vec vb)\}),
\label{Glileitrans}
\end{equation}
and furnishes a ray-representation whose multiplier exponents
are given by (\ref{Gphase}). It is easy to see that the multiplier 
exponents are non-trivial, i.e., not removable by a redefinition 
(\ref{redef}). The quickest way to see this is as follows: suppose 
to the contrary that they were trivial and that hence (\ref{redef}) 
holds with $\xi'\equiv 0$. Trivially, this equation will continue to 
hold after restriction to any subgroup $G_0\subset G$. We choose for 
$G_0$ the abelian subgroup generated by boosts and space translations,
so that the combination $\gamma(g_1)-\gamma(g_1g_2)+\gamma(g_2)$
becomes symmetric in $g_1,g_2\in G_0$. But the exponent (\ref{Gphase})
stays obviously asymmetric after restriction to $G_0$. Hence no 
cancellation can take place, which contradicts our initial assumption.

The same trick immediately shows that the multiplier exponents are
inequivalent for different total masses $M$. Hence, by the general 
argument given in the previous chapter, if $\H'$ and $\H''$
correspond to Hilbert spaces of states with different overall masses 
$M'$ and $M''$, then the requirement that the Galilei group should act 
on the set of physical states excludes superpositions of states of 
different overall mass. This is Bargmann's superselection rule.

I criticize these arguments for the following reason: The dynamical 
framework that we consider here treats `mass' as parameter(s) which 
serves to specify the system. States for different overall masses are 
states of {\it different} dynamical systems, to which the superposition 
principle does not even potentially apply. In order to investigate a 
possible violation of the superposition principle, we must find a 
dynamical framework in which states of different overall mass are states 
of the {\it same} system; in other words, where mass is a dynamical  
variable. But if we enlarge our system to one where 
mass is dynamical, it is not at all obvious that the Galilei group 
will survive as symmetry group. We will now see that in fact it 
does not, at least for the simple dynamical extension which we
now discuss.

The most simple extension of the classical model is to maintain the 
Hamiltonian, but now regarded as function on an extended,  
$6n+2n$ - dimensional phase space with extra `momenta' $m_i$ and
conjugate generalized `positions' $\lambda_i$. Since the 
$\lambda_i$'s do not appear in the Hamiltonian, the $m_i$'s are
constants of motion. Hence the equations of motion for the 
$\vec x_i$'s and their conjugate momenta $\vec p_i$ are unchanged 
(upon inserting the integration constants $m_i$) and those of the 
new positions $\lambda_i$ are 
\begin{equation}
\dot\lambda_i(t)=\frac{\partial V}{\partial m_i}-\frac{\vec p_i^2}{2m_i^2},
\label{zetadot}
\end{equation}
which, upon inserting the solutions $\{\vec x_i(t),\vec p_i(t)\}$,
are solved by quadrature. 

Now, the point is that the new Hamiltonian equations of motion do not 
allow the Galilei group as symmetries anymore. But they do allow the 
$\reals$-extension $\b G$ as symmetries \cite{Giulini2}. Its 
multiplication law is given by (\ref{extension}), with $\xi$  as in 
(\ref{Gphase}). The action of $\b G$ on the extended space of
$\{\hbox{configurations}\}\times\{\hbox{times}\}$ is now given by 
\begin{eqnarray}
&&\b g(\{\vec x_i\},\{\lambda_i\},t)=
(\theta,R,\vec v,\vec a,b)(\{\vec x_i\},\{\lambda_i\},t)\nonumber\\
&&=(\{R\vec x_i+\vec v t+\vec a\}\,,\,
\{\lambda_i-(\hbarM\theta+\vec v\cdot R\cdot\vec x_i+\shalf\vec v^2t)\}
\,,\, t+b).
\label{newaction}
\end{eqnarray}
With (\ref{extension}) and (\ref{Gphase}) it is easy to verify that 
this defines indeed an action. Hence it also defines an action on 
curves in the new Hilbert space $\b\H:=L^2(R^{4n},d^{3n}\vec x d^n\lambda)$, 
given by 
\begin{equation}
{\b{\op T}}_{\b g}\psi:=\psi\circ {\b g}^{-1}\,,
\label{newtrans}
\end{equation}
which already maps solutions of the new Schr\"odinger equation to
solutions, without invoking non-trivial phase factors. This simple
transformation law contains the more complicated one (\ref{trans}) 
upon writing $\b\H$ as a direct integral of vector spaces 
$\H_{\{m_i\}}$, each isomorphic to our old $\H$. Then, for each 
$n$-tuple of masses $\{m_i\}$, the new Schr\"odinger equation 
reduces to the old one on $\H_{\{m_i\}}$ and (\ref{newtrans}) 
restricts to (\ref{trans}) \cite{Giulini2}. 

In the new framework the overall mass, $M$, is a dynamical variable, 
and it would make sense to state a superselection rule with  respect 
to it. But now $\b G$ rather than $G$ is the dynamical symmetry group, 
which acts by a proper unitary representation on $\b\H$, so that the  
requirement that the dynamical symmetry group should act on the space 
of physical states will now not lead to any superselection rule. Rather, 
the new and more physical interpretation of a possible superselection 
rule for $M$ would be that we cannot localize the system in the 
coordinate conjugate to overall mass, which we call 
$\Lambda$, i.e., that only the {\it relative} new positions 
$\lambda_i-\lambda_j$ are observable.\footnote{A  system 
$\{(\t\lambda_i,\t m_i\})$ of canonical coordinates including 
$M=\sum_im_i$ is e.g. $\t\lambda_1:=\lambda_1$, ${\t m}_1=M$ and 
$\t\lambda_i=\lambda_i-\lambda_1$, ${\t m}_i=m_i$ for $i=2...n$. Then 
$\Lambda=\t\lambda_1$.}
(This is so because $M$ generates translations of equal amount in all 
$\lambda_i$.) But this would now be a contingent physical property rather 
than a mathematical necessity. Note also that in our 
dynamical setup it is inconsistent to just state that $M$ generates 
gauge symmetries, i.e. that $\Lambda$ corresponds to a physically non 
existent degree of freedom. For example, a motion in real time
along $\Lambda$ requires a non-vanishing action (for non-vanishing 
$M$), due to the term $\int dt\, M\dot\Lambda$ in the expression for 
the action. 

If decoherence were to explain the (ficticious) mass superselection
rule, it  would be due to a dynamical instability (as explained
in \cite{Joos}) of those states which are more or less localized in
$\Lambda$. Mathematically this effect would be modelled by removing 
the projectors onto  $\Lambda$-subintervalls from the algebra of 
observables, thereby putting $M$ (i.e. its projectors) into the 
center of $\O$. Such a non-trivial center should therefore be 
thought of as resulting from an approximation-dependent idealisation.

\section{Charge Superselection Rule}
In the previous case I said that superselection rules should be
stated within a dynamical framework including  as dynamical
degree of freedom the direction generated by the superselected 
quantity. What is this degree of freedom in the case of a
superselected electric charge and how does it naturally appear
within the dynamical setup? What is its relation to the 
Coulomb field whose r\^ole in charge-decoherence has been suggested 
in \cite{GKZ}? In the following discussion I wish to investigate 
into these questions by looking at the Hamiltonian formulation of 
Maxwell's equation and the associated canonical quantization. 

In Minkowski space, with preferred coordinates $\{x^{\mu}=(t,x,y,z)\}$ 
(laboratory rest frame), we consider the spatially finite region 
$Z=\{(t,x,y,z):\, x^2+y^2+z^2\leq R^2\}$. $\Sigma$ denotes the intersection 
of $Z$ with a slice $t=\hbox{const.}$ and $\partial\Sigma=:S_R$ its 
boundary (the laboratory walls). Suppose we wish to solve Maxwell's equations 
within $Z$, allowing for charged solutions. It is well known that in order 
for charged configurations to be stationary points of the action, the 
standard action functional has to be supplemented by certain surface 
terms (see e.g. \cite{Gervais-Zwanziger}) which involve new fields
on the boundary, which we call $\lambda$ and $f$, and which represent 
a pair of canonically conjugate variables in the Hamiltonian sense.
On the laboratory walls, $\partial\Sigma$, we put the boundary 
conditions that the normal component of the current and the
tangential components of the magnetic field vanish. Then the appropriate 
boundary term for the action reads
\begin{equation}
\int_Z dt\, d\omega (\dot\lambda+\phi)f,
\label{boundary}
\end{equation}
where $\phi$ is the scalar potential and $d\omega$ the measure on the 
spatial boundary 2-sphere rescaled to unit radius.
Adding this to the standard action functional and expressing all 
fields on the spatial boundary by their multipole moments (so that 
integrals $\int_{\partial\Sigma} d\omega$ become $\sum_{lm}$),
one arrives at a Hamiltonian function
\begin{equation}
H=\int_{\Sigma}\left[\shalf(\vec E^2+(\vec\nabla\times\vec A)^2)+
\phi(\rho-\vec\nabla\cdot\vec E)-\vec A\cdot\vec j\right]
+\sum_{lm}\phi_{lm}(E_{lm}-f_{lm}).
\label{Hamiltonian}
\end{equation}
Here the pairs of canonically conjugate  variables are 
$(\vec A(\vec x),-\vec E(\vec x))$ and $(\lambda_{lm},f_{lm})$, 
and $E_{lm}$ are the multipole components of
$\vec n\cdot\vec E$, where $\vec n$ is the normal to $\partial\Sigma$.
The scalar potential $\phi$ has to be considered as Lagrange multiplier. 
With the given boundary conditions the Hamiltonian is differentiable 
with respect to all the canonical variables\footnote{This would not 
be true without the additional surface term (\ref{boundary}). 
Without it one does not simply obtain the wrong Hamiltonian 
equations of motions, but none at all! Concerning the Langrangean
formalism one should be aware that the Euler-Lagrange equations 
may formally admit solutions (e.g. with long-ranged (charged) 
fields) which are outside the class of functions which one used 
in the variational principle of the action (e.g. rapid fall-off).  
Such solutions are not stationary points of the action and their 
admittance is in conflict with the variational principle unless 
the expression for the action is modified by appropriate boundary 
terms.}
and leads to the following  equations of motion
\begin{eqnarray}
\dot{\vec A} 
&=& \frac{\delta H}{\delta (-\vec E)}
=-\vec E-\vec\nabla\phi\,,                               
\label{motion1}                                       \\
-\dot{\vec E} 
&=&-\frac{\delta H}{\delta\vec A}
=\vec j-\vec\nabla\times\vec (\vec\nabla\times\vec A)\,,
\label{motion2}                                       \\
{\dot\lambda}_{lm} 
&=& \frac{\partial H}{\partial f_{lm}}=-\phi_{lm}\,,     
\label{motion3}                                       \\
{\dot f}_{lm}
&=& -\frac{\partial H}{\partial\lambda_{lm}}=0\,.        
\label{motion4}                                       
\end{eqnarray}
These are supplemented by the equations which one obtains by varying 
with respect to the scalar potential $\phi$, which, as already said, 
is considered as Lagrange multiplier. Varying first with respect to
$\phi(\vec x)$ (i.e. within $\Sigma$) and then with respect to 
$\phi_{lm}$ (i.e. on the boundary $\partial\Sigma$), one obtains

\begin{eqnarray}
G(\vec x):&=&\vec\nabla\cdot\vec E(\vec x)-\rho(\vec x)=0,
\label{constraint1}                                      \\
G_{lm}:&=& E_{lm}-f_{lm}=0.
\label{constraint2}                                      
\end{eqnarray}
These equations are constraints (containing no time derivatives) which, 
once imposed on initial conditions, continue to hold due to the 
equations of motion.\footnote{Equation (\ref{motion2}) together with 
charge conservation, $\dot\rho+\vec\nabla\cdot\vec j=0$, shows that 
(\ref{constraint1}) is preserved in time, and (\ref{motion4}) together 
with the boundary condition that $\vec n\cdot \vec j$ and 
$\vec n\times(\vec\nabla\times\vec A)$  vanish on $\partial\Sigma$
show that (\ref{constraint2}) is preserved in time.} 

This ends our discussion of the classical dynamical theory. The point 
was to show that it leaves no ambiguity as to what its dynamical degrees 
of freedom are, and that we had to include the variables $\lambda_{lm}$
along with their conjugate momenta $f_{lm}$ in order to gain consistency 
with the existence of charged configurations. The physical interpretation 
of the $\lambda_{lm}$'s  is not obvious. Equation (\ref{motion3}) merely 
relates their time derivative to the scalar potential's multipole moments 
on the boundary, which are clearly highly non-local quantities. The 
interpretation of the $f_{lm}$'s follow from (\ref{constraint2})
and the definition of $E_{lm}$, i.e. they are the multipole moments 
of the local charge-flux $\vec n\cdot \vec E$. In particular, 
for $l=0=m$ we have 
\begin{equation}
f_{00}=(4\pi)^{-\frac{1}{2}}\, Q,  
\label{charge}
\end{equation}
where $Q$ is the total charge of the system. Hence we see that the total 
charge  generates motions in $\lambda_{00}$. But this means that the 
degree of freedom labelled by $\lambda_{00}$ truly exists (in the
sense of the theory). For example, a motion along $\lambda_{00}$ will
cost a non-vanishing amount of action 
$\propto Q(\lambda_{00}^{\rm final}-\lambda_{00}^{\rm initial})$. 
A declaration that $\lambda_{00}$ really labels only a gauge degree
of freedom is {\it incompatible} with the inclusion of charges states.
Similar considerations apply of course to the other values of $l,m$. 
But note that this conclusion is independent of the radius $R$ of the 
spatial boundary 2-sphere $\partial\Sigma$. In particular, it 
continues to hold in the limit $R\rightarrow\infty$. We will not 
consistently get rid of physical degrees of freedom that way, even if we 
agree that realistic physical measurements will only detect field values 
in bounded regions of space-time. See \cite{Giulini3} for more 
discussion on this point and the distinction between proper 
symmetries and gauge symmetries. 

It should be obvious how these last remarks apply to the statement of a 
charge superselection rule. Without entering the technical issues 
(see e.g. \cite{Strocchi-Wightman}), its basic ingredient is Gauss' law
(for operator-valued quantities), locality of the electric field and 
causality. That $Q$ commutes with all (quasi-) local observables then 
follows simply from writing $Q$ as surface integral of the local flux 
operator $\vec n\cdot\hat{\vec E}$, and the observation that the surface 
may be taken to lie in the causal complement of any bounded space-time 
region. Causality then implies commutativity with any local observable.

In a heuristic Schr\"odinger picture formulation of QED one 
represents states $\Psi$ by functions of the configuration 
variables $\vec A(\vec x)$ and $\lambda_{lm}$. The momentum operators  
are obtained as usual: 
\begin{eqnarray}
-\vec E(\vec x)&\longrightarrow & 
-{\rm i}\frac{\delta}{\delta\vec A(\vec x)}\,, 
\label{quant1}\\
f_{lm}&\longrightarrow & -{\rm i}\frac{\partial}{\partial\lambda_{lm}}\,.
\label{quant2}
\end{eqnarray}
In particular, the constraint 
(\ref{constraint2}) implies the statement that on physical 
states $\Psi$ we have\footnote{Clearly all sorts of points are simply 
sketched over here. For example, charge quantization presumably means 
that $\lambda_{00}$ should be taken with a compact range, which in turn 
will modify (\ref{quant2}) and (\ref{Psi}). But this is irrelevant to 
the point stressed here.}
\begin{equation}
\hat Q\Psi=-{\rm i}\sqrt{4\pi}
\frac{\partial}{\partial\lambda_{00}}\Psi\,.
\label{Psi}
\end{equation}
This shows that a charge superselection rule is equivalent to the
statement that we cannot localize the system in its $\lambda_{00}$
degree of freedom. Removing {\it by hand} the multiplication
operator 
$\lambda_{00}$ (i.e. the  projectors onto $\lambda_{00}$-intervals) 
from our observables clearly makes $Q$ a central element in the 
remaining algebra of observables. But what is the physical justification 
for this removal? Certainly, it is valid FAPP if one restricts to local 
observations in space-time. To state that this is a {\it fundamental} 
restriction, and not only an approximate one, is equivalent to saying 
that for some fundamental reason we cannot have access to some 
of the {\it existing} degrees of freedom, which seems at odds with 
the dynamical setup. Rather, there should be a 
dynamical reason for why localizations in $\lambda_{00}$ seem FAPP
out of reach. Again, if decoherence is to explain the phenomena 
of an apparent charge superselection rule, localization in 
$\lambda_{00}$ must be highly unstable against dynamical 
decoherence.

Regarding the charge superselection rule, Erich~Joos asked the 
following question in section 3.2 of his contribution to this volume: 
``What is the {\it quantum} physical r\^ole of the Coulomb
field''. In some sense the analysis given here is also meant to shed
some light on this question. We have seen the necessity to 
include the non-local canonical variables $\lambda_{lm},f_{lm}$ 
and discussed their relation with the $\frac{1}{r^2}$ - part  
of the electric field. Clearly, the arguments given here are 
neither complete nor rigorous in any sense. What they suggest 
is to relate the $\lambda_{lm}$  degrees of freedom to the 
precise infrared structure of QED, for example along the lines of 
\cite{Zwanziger1}\cite{Zwanziger2}\cite{Gervais-Zwanziger}.
Eventually this raises the question of how to fully describe 
the state space of QED, which is known to be a notoriously 
difficult problem~\cite{Buchholz}. 

Also, from my presentation 
it did not become apparent why charge plays such a particular
r\^ole among the other superselection operators ${\hat f}_{lm}$ 
associated with the higher multipole moments of the asymptotic 
flux distribution. This question is further analyzed in 
\cite{Buchholz} and from a more concrete lattice-calculational
point of view in \cite{Kijowski}.

\vfill\eject

\end{document}